\documentclass[preprint,aps,amsmath,showkeys,showpacs,floatfix]{revtex4}
\usepackage{graphicx}% Include figure files
\usepackage{dcolumn}% Align table columns on decimal point
\usepackage{bm}% bold math

\begin{document}

% *********************************************************
%   linea anyadida para que reduzca espacio entre lineas
% *******************************************************
%\tightenlines
% ********************************************************

\title{Collective excitations in double quantum dots}
\author{N. Barber\'an}
\affiliation{Departament d'Estructura i Constituents de la
Mat\`eria,
Facultat de F\'\i sica, Universitat de Barcelona,
E-08028 Barcelona, Spain}

\vskip4mm

\begin{abstract}
We analyse the ground state properties of vertical-double quantum
dots
in the lowest Landau level regime for filling factor $\nu=2$. This
analysis follows two lines: on the one hand, we study the
dispersion relation of different collective excitations and look for
roton-type minima that signal some type of order
within the ground state. On the other hand, we calculate expected
values of the antiferromagnetic order operator defined as
$\langle S_x^RS_x^L \rangle$. We conclude that of the three
existing types of ground state at $\nu=2$ the canted and symmetric
states show spin order on the plane of the dots. In the case of the
canted state,
two types of spin order are shown as a
consequence of the coupling between spin and isospin. We propose that
one of the excitations obtained over the symmetric state could be the
origin of one of the decoherence processes detected experimentally in
qubit systems \cite{hay}.
\end{abstract}
\vskip2mm

\pacs{73.21.La, 73.43.Lp.}
\keywords{Collective excitations, double quantum dot,
incompressible states, exact diagonalization.}
\maketitle
\vfill
\eject

\subsection*{I. Introduction}

The possibility of tunneling between the left and right layers in a
double-layered system introduces an extra energy scale added to those
typical of a system defined by a quantum well in a perpendicular
magnetic field, namely, the Coulomb and Zeeman energies (under the
assumption of
the lowest Landau level (LLL) regime
the kinetic energy is frozen and can be ignored). This
extra degree of freedom yields a rich variety of ground
states (GS) as well as different excited configurations. In a previous
work Das Salma et al. \cite{das} established that for filling factor
$\nu=2$ ( one for each layer) there is a universal GS phase diagram
that contains only three possibilities: the ferromagnetic
state $\mid FER \rangle$ with all spins up and equal populations of
symmetric and antisymmetric states (the symmetric and antisymmetric
combinations of the left and right single particle states), the
symmetric state $\mid SYM \rangle$ with all the electrons in the
symmetric state and the spin up and down equally populated and, in
between, the canted state $\mid CAN \rangle$ characterized by a
mixture
of spins and isospins. The isospin (or "pseudospin") is the quantum
number associated with the symmetric and antisymmetric single particle
states. The balance between the Zeeman energy, the tunneling strength
and the Coulomb interaction determines the boundaries within the phase
diagram. Of special interest is the canted state
characterized by ferromagnetic order in the perpendicular direction
and antiferromagnetic order (AFO) on the plane of the layers. This
order
in plane is a consequence of the broken symmetry produced by the
degeneracy of the GS. The excited state produced by the operator
that changes spin and isospin in one unit and leaves the total angular
momentum unchanged is degenerated with the GS. This degeneracy is
produced in spite of the fact that the Zeeman and tunneling gaps (
$\Delta_Z$ and $\Delta_{t}$ respectively) are
non zero, a necessary condition to obtain canted states.
The AFO in plane defined as:
\begin{equation}
\langle S_x^R \rangle\,\,-\,\,\langle S_x^L \rangle
\end{equation}
is, within a Hartree-Fock (HF) calculation, different from zero only
for the canted state \cite{das}.

\medskip

The aim of this report is to analyse to what extent this
phenomenology is preserved for a finite system built up from
a vertical double quantum dot (DQD) confined by identical parabollic
potentials in each layer, preserving the conditions of filling factor
$\nu=2$ (for an even number of electrons) and LLL regime.

\medskip

This paper is organized as follows: in Sec.II we define the
lines chosen to analyse the properties of the GS configurations,
namely, the study of the collective mode excitations and the
calculation of the expected value of an appropriate
AFO operator.
In Sec.III we concentrate on the analysis of the
collective modes and on the information about the GS that derives
from its properties. In Sec.IV the results obtained from the
AFO operator are shown, and finally in Sec.V we
draw our conclusions.

\medskip

\subsection*{II. Ground state configurations}

In this Section we address the comparison between a DQD
finite system of $N$ electrons and the double layer with which it
shares
the same number of degrees of freedom. It should be noted that
the question related to the type of GS's has been solved
in Ref.\cite{tej} within an exact diagonalization calculation
(EDC). It
has been proved that, similarly to the double layered case, the states
$\mid FER\rangle$, $\mid SYM\rangle$ and $\mid CAN\rangle$ cover all
the possibilities. The only difference is that there are several
possible canted states, the number of which grows with the number of
electrons. For $N=6$ for instance, and $\nu=2$, the total spin
ranges from
$S_z=3$ for the $\mid FER\rangle$ state to zero for the $\mid
SYM\rangle$ state and in between there are two canted states with
total spins $S_z=2$ and $1$ respectively. At the same time the parity
($1$ or $-1$) of the GS changes simultaneously with each spin
transition. Parity is the well defined quantum number related to
specular symmetry between the dots and is defined as $P=(-1)^{X/2}$,
where $X=N_S-S_A$ is the balance between symmetric and antisymmetric
single particle occupied states \cite{tej}.
It should be pointed out that neither the single particle layer
occupation states $\mid R\rangle$ or $\mid L\rangle$  nor the
symmetric or antisymmetric combinations $\mid S\rangle$ or
$\mid A\rangle$ are associated with
well defined single particle quantum numbers \cite{bar}; in other
words
the Coulomb interaction mixes isospin meanwhile tunnelig mixes layer
occupations.

\medskip

In this report, we examine the
structure of the different GS's. On the one hand, we
take advantage of the finitness of the system and perform EDC,
which gives exhaustive information of
the spectrum. However, since the diagonalization is
performed within separate subspaces of well defined $M$ (total angular
momentum), $S_z$ (total spin) and $P$ (parity) defining a
configuration denoted by $(M,S_z,P)$, no information can be obtained
from the AFO in plane defined by Eq.(1) since the
expected values vanish.
In order to analyse the presence of some type of order in the GS
we follow two lines:

\medskip

The first line is based on the results
of the analysis of the dispersion relation of the
density wave excitations of liquid helium. On the one hand
the
roton minimum gives indirect information about the short-range order
between helium atoms in the GS and on the other hand, since a single
mode approximation accurately reproduces the experimental
data for small transferred momentum \cite{jac},
it shows that a collective excitation built up from a linear
combination of single-body excitations provides a suitable model
within this region of the dispersion relation.
The most convenient excitations to analyse the GS properties are
those that change spin and parity as is
discussed in the next section.

\medskip

The second line is the search for an appropriate operator
which reveals the presence of order in the plane of the dots and
has non-vanishing
expected values for well defined configurations. To this end we
chose the operator given by:
\begin{equation}
S_x^R S_x^L\,\,\,\,\,,
\end{equation}
which complements the information provided by the first line.

\medskip

\subsection*{III. Collective excitations}

In order to choose the appopriate excitations, we considered one
of the
results obtained in a previous paper \cite{bar} for the same system.
Among all the possibilities to excite the system, the excitation that
simultaneously changes spin and parity in one unit and leaves $M$
unchanged has some important properties: it is the absolute lowest
energy excitation over the three types of GS and softens at the
boundaries of the phase diagram ( for $\Delta_Z$ over $\Delta_{t}$)
signaling the frontiers within the diagram.
By the increase of $\Delta_{t}$, for fixed $\Delta_Z$, the GS
for $N=6$ evolves as
$\mid FER\rangle
\rightarrow \mid CAN1 \rangle \rightarrow \mid CAN2\rangle
\rightarrow \mid SYM\rangle$.
In Fig.1 we show the
evolution of the excitation energy over the GS's ( $(6,3,-1)$,
$(6,2,1)$, $(6,1,-1)$ and $(6,0,1)$) as
$\Delta_{t}$ changes. It should be noted that over the canted states,
there are two different possibilities, namely, $(6,2,1)\rightarrow
(6,3,-1)$ or $(6,2,1)\rightarrow(6,1,-1)$. The transitions take place
at
the values of $\Delta_{t}$ where the excitation energy goes to zero.
This led us to study the dispersion
relation of this type of collective mode as a function of the
transferred angular momentum
($\Delta M=l$). We will denote these excitations as $SIE$
(spin-isospin-excitation). Furthermore, in order to complete the
scenario,
we also analysed the evolution with $l$ of three other possible
excitations: the charge
density wave (CDW) for which the spin and parity remain constant, the
one that leaves the parity unchanged (SE) and the one that leaves the
spin unchanged (IE). The results are shown in Figs.2-4 for a DQD of
six electrons. Similar phenomenology was found for $N=8$.
Even though no special structure
is shown for the $\mid FER\rangle$ state as  expected, some
minima appear over the $\mid CAN1\rangle$ and over the $\mid
SYM\rangle$ states whenever the spin
change is involved. For the $\mid CAN1\rangle$ state the presence or
absence of parity change modifies
the value of $l_c$ on which the local minimum occurs, namely,
spin
and isospin appear to be coupled within the canted state, or in other
words, two different spin ordered states seem to coexist in a nearly
degenerate state.
It must be emphasized that the energy gap
between
the GS and the SIE for $l=0$ is negligible as compared to the typical
excitation energies of the SIE for $l\neq 0$, as can be inferred from
the comparison of the energy scales in Figs.1
and 2-4. The degeneracy between the GS and the two SIE would be
complete for large number of electrons \cite{das}.
For the $\mid
SYM\rangle$ state there is a strong minimum in the
SE spectrum and no structure is apparent in the SIE.
The previous result suggests
that there is spin order in the $\mid SYM\rangle$ state
in contrast to the conclusions inferred from
the results obtained by the use of Eq.1 in a HF calculation for double
layer systems \cite{das}. This interpretation
is reinforced by the results obtained by the use of Eq.2
as is discussed in the next Section.

%%%%%%%%%%%%%%%%%%%%%%%%%%%%%%%%%%%%%%%%%%%%%%%%%%%%%%%%%%%%%%%

\begin{figure}[htb]
\includegraphics*[width=0.7\columnwidth]{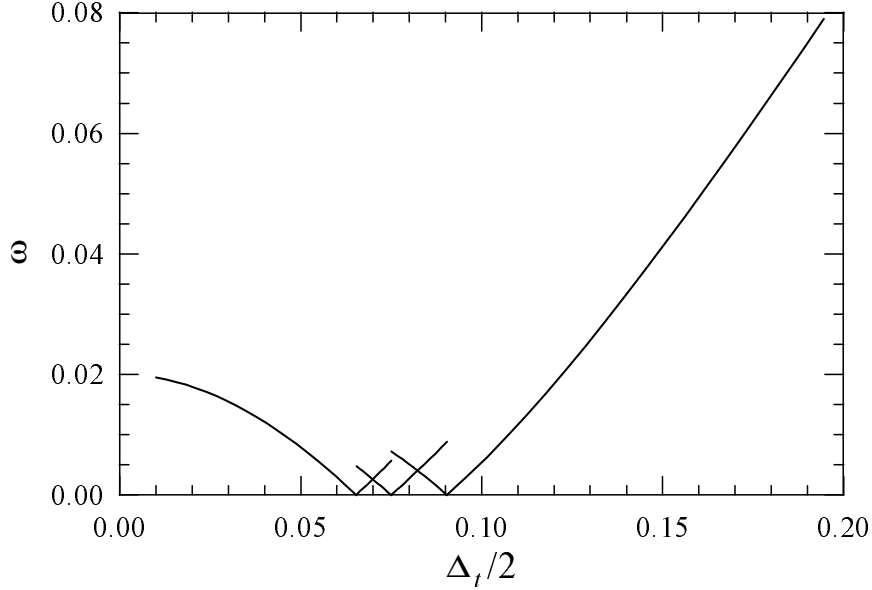}
\caption{Lowest energy excitation for $N=6$ as a function of
$\Delta_t$ for different regions of the GS phase diagram. Energies
are given in units of $e^2/\epsilon l_B$ where $l_B$ is the magnetic
length and $e$ and $\epsilon$ are the electronic charge and the
dielectric constant of the host semiconductor respectively. The
excited configurations are (from left to right) $(6,3,1)$, $(6,2,1)$
and $(6,1,-1)$ for the positive-gradient curves and $(6,2,1)$,
$(6,1,-1)$ and $(6,0,1)$ for the negative-gradient curves.}
\end{figure}

%%%%%%%%%%%%%%%%%%%%%%%%%%%%%%%%%%%%%%%%%%%%%%%%%%%%%%%%%%%%%%%%
\medskip

Other interesting properties derive from the analysis of the
collective modes. On the one hand, the local minima at $l_c$ are
characterized by states consisting of only one Slater determinant,
(the spectral weight of the most important Slater within state at
$l_c$ is
given by $0.96$ for the SE and by $0.98$ for the SIE over the $\mid
CAN1\rangle$
state). In other words, the excitations at $l_c$ are non-correlated
states for which
the system develops a strong expansion. This expansion is a direct
consequence of the restrictions arising from the changes imposed
over spin
and parity. For instance, for the $\mid
CAN1\rangle$ GS configuration  $(6,2,1)$, the SIE is $(6+l,3,-1)$
and the SE is $(6+l,3,1)$, while
$P=-1$ is obtained
from $X=-4,0$ or $4$ and $P=1$ from $X=-6,-2,2$ or $6$ in such a way
that the expansions shown in the SIE from $l=3$ to $4$ or from
$l=8$
to $9$ in the SE correspond to changes of $X$ from $0$ to $4$ and
from
$2$ to $6$ respectively. Therefore, the increase of $l$ produces a
sudden expansion in the excited structure since the new allowed values
of
$X$ are compatible with the occupation of larger single-particle
angular momenta.
The strong expansion reduces the Coulomb interaction and
produces the minima. We emphasize that the same effect is
expected at some $l$ for the dispersion relation of each excitation,
although the expansion produces a strong bound state only for those
signaled by $l_c$ in Figs.2-4.

%%%%%%%%%%%%%%%%%%%%%%%%%%%%%%%%%%%%%%%%%%%%%%%%%%%%%%%%%%%%

\begin{figure}[htb]
\includegraphics*[width=0.7\columnwidth]{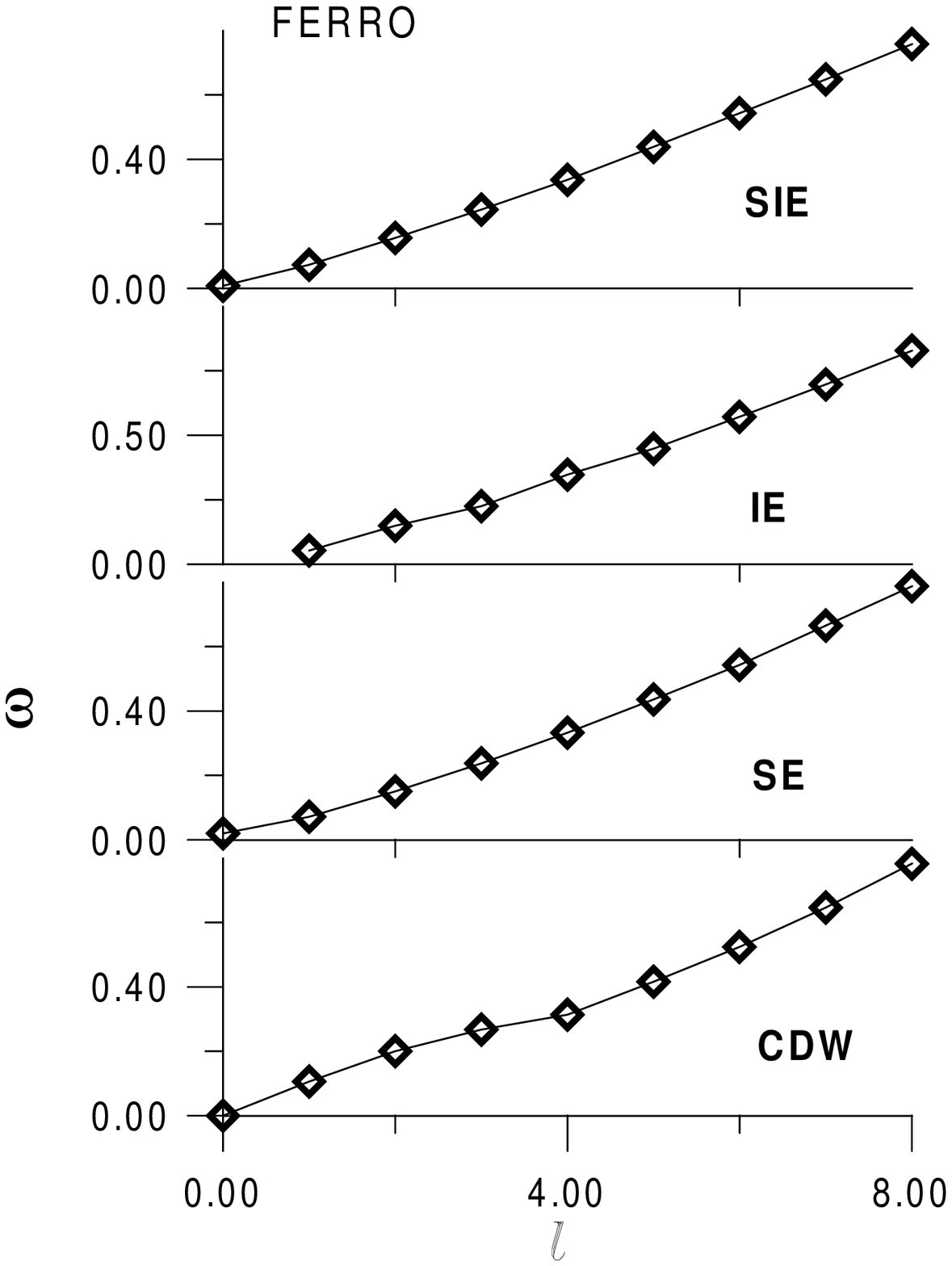}
\caption{Dispersion relations for different excitations over the $\mid
FER\rangle$ GS (see text), calculated at $\Delta_t=0.05$. The same
units as in Fig.1 are considered.}
\end{figure}

\begin{figure}[htb]
\includegraphics*[width=0.7\columnwidth]{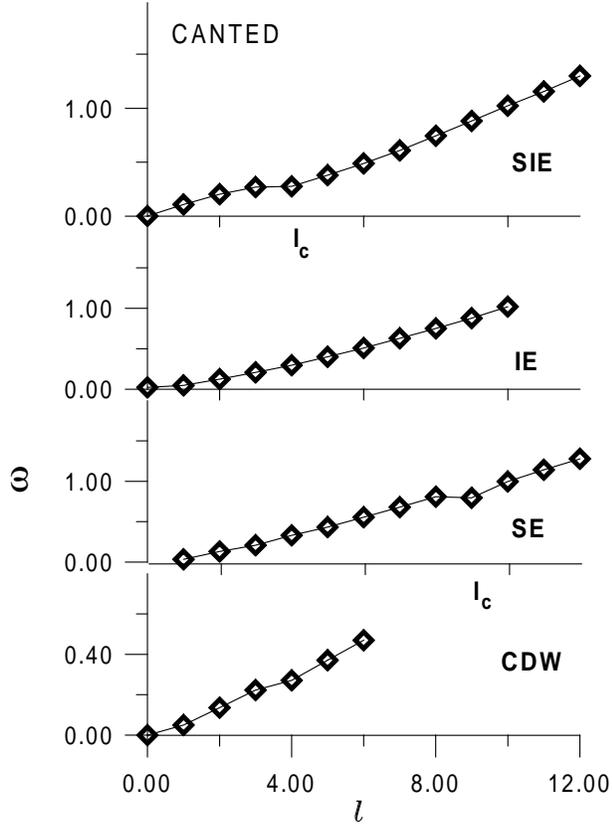}
\caption{The same as Fig.2 for the $\mid CAN1\rangle$ GS calculated at
$\Delta_t=0.07$. The critical transferred angular momenta at the
minima are signaled by $l_c$.}
\end{figure}

\begin{figure}[htb]
\includegraphics*[width=0.7\columnwidth]{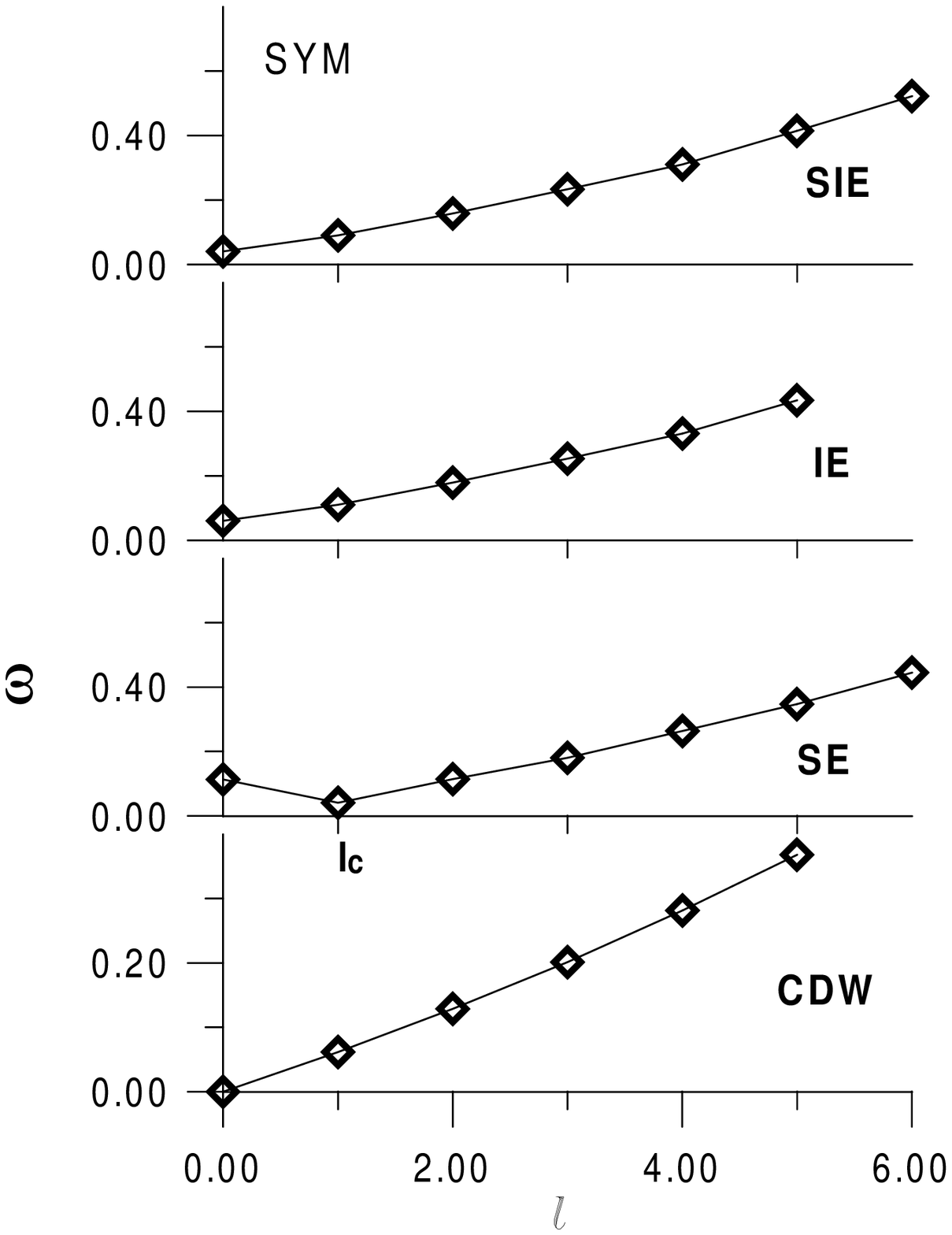}
\caption{The same as Fig.3 for the $\mid SYM\rangle$ GS calculated at
$\Delta_t=0.15$.}
\end{figure}

%%%%%%%%%%%%%%%%%%%%%%%%%%%%%%%%%%%%%%%%%%%%%%%%%%%%%%%%%%%%%%%%

\medskip

Furthermore, the states at $l_c$ cannot be modelled within a single
mode
approximation. This can be confirmed directly from the diagonalization
output and from the fact that the excited states contain effectively
only one Slater. The GS $(6,2,1)$ is a combination of 18 Slater
determinants, although only one of them ( with negligible
spectral weight) reproduces
the excited SIE configuration by a one body excitation, and
non of them
reproduces the SE configuration.
Two or more body excitations are necessary to approach the
excited configuration. In contrast, the $\mid CAN1\rangle$ GS can
be reproduced by
\begin{equation}
Op\mid FER\rangle=\Psi
\end{equation}
where
\begin{equation}
Op=\sum_l(c^+_{l\downarrow S}\,\,c_{l\uparrow
A}\,\,+\,\,c^+_{l\downarrow A}\,\,c_{l\uparrow S})
\end{equation}
which has for $8$ electrons, an overlap (i.e., $\mid\langle\Psi\mid
CAN1\rangle \mid^2$) of $0.943$. Moreover, as the $\mid CAN1\rangle$
state can
be interpreted as an excitation of the $\mid FER\rangle$ GS, it proves
that the excitation for $l=0$ in Fig.2 for the SIE can be
reproduced
with a single mode approximation, as had to be the case for long
wavelength excitations.

\medskip

\subsection*{IV. Antiferromagnetic order in plane}

In order to obtain independent information about the spin
configuration in plane, we looked for an order operator with expected
values that were not necessarily zero for well defined configurations.
Fig.5
shows the expectation values of the operator given by Eq.2 as a
function of $\Delta_{t}$. The $\mid FER\rangle$ state has
zero antiferromagnetic order as expected; however,
the $\mid SYM\rangle$ as well as the $\mid CAN1\rangle$ and $\mid
CAN2\rangle$ states show non-vanishing values.

%%%%%%%%%%%%%%%%%%%%%%%%%%%%%%%%%%%%%%%%%%%%%%%%%%%%%%%%%%%%%%%

\begin{figure}[htb]
\includegraphics*[width=0.7\columnwidth]{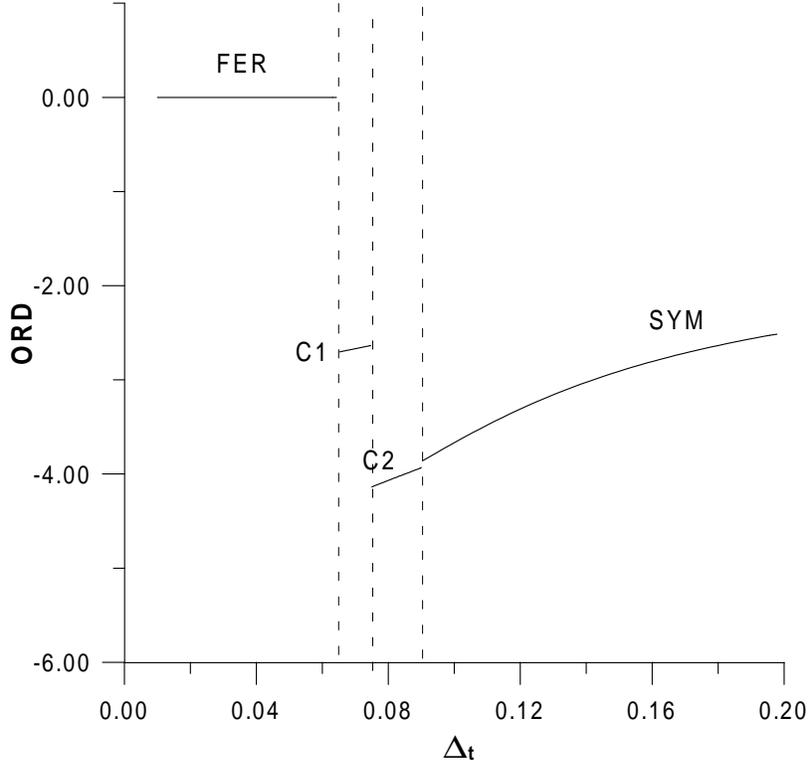}
\caption{The order parameter defined as $ORD=\langle
S_x^RS_x^L\rangle$ calculated for the different GS's over the phase
diagram as a function of $\Delta_t$.}
\end{figure}

%%%%%%%%%%%%%%%%%%%%%%%%%%%%%%%%%%%%%%%%%%%%%%%%%%%%%%%%%%%%%%%%

\medskip

In fact, there is no
any contradiction between our results and those given in
Ref.\cite{das} as regards the $\mid SYM\rangle$ state. The
following situation is possible, namely, that the expected value of
$S_x^R$
( or
$S_x^L$) vanishes as an integrated quantity over the plane of the
layer though the expected value of the local operator
$S_x^R(x,y)$ does not vanish,
in such a way that $\langle S_x^RS_x^L\rangle \neq 0$ and
$\langle S_x^R\rangle = \langle S_x^L\rangle =0$ even within a
Hartree-Fock calculation in which the GS is a mixture of
configurations.

\medskip

\subsection*{V. Conclusions and discussion}

The observation of minima in the excitation
spectrum of special excited configurations $(M+l,S_z\pm1, \pm P)$
indicates spin order in the GS. This result is in
accordance with the results obtained from the expectation value of
the AFO in plane operator defined as
$\langle S_x^R S_x^L\rangle$.

\medskip

In contrast with the findings obtained
for
double layers within a HF approximation and using
Eq.1, the $\mid SYM\rangle$ state has spin
order in  the plane of the dots as defined by Eq.2. However, as even
within a HF calculation the
zero value of Eq.1 could hide local AFO, the difference between our
findings for a DQD and those of Ref.\cite{das} for double layers
cannot be imputed to finite size effects without deeper analysis.
For the $\mid CAN\rangle$ state, as
mentioned above,
the excitations at $l_c$ need two or more body
operators to be generated from the GS. Meanwhile, for the $\mid
SYM\rangle$
state, the excitation at $l_c$ can be accurately
reproduced from a single-body operator.

\medskip

For all cases, the excited states at the local minima are
non-correlated states
of a strongly expanded configuration as compared to the GS or the
excited state at $l_c-1$.
The parameter $l_c$ is associated with a length
scale that characterizes
spin and isospin order in the $\mid CAN\rangle$
state and spin order in the $\mid SYM\rangle$ state.

\medskip

As a last remark, increasing interest has
recently been shown in two-level systems
which can be used as qubits in quantum information technology and
particularly in coupled quantum dots. In a recent peper \cite{hay}
Hayashi et al. report coherent oscillations ( in qubits built from
lateral DQD's ) observed for several combinations of ground
and excited states. The coherent oscillations between left and right
localized
states are obtained for zero offset when the tunneling gap is the most
important energy
scale, in such a way that the system is effectively in the $\mid
SYM\rangle$ state.
For this state, the understanding of the
possible decoherent processes is crucial for practical applications.
The excited state found at
$l_c$ in Fig.4  may be involved in one of the decoherent mechanisms
since it is
the lowest energy excitation and in this case, the decoherence
would be characterized by a spin flip process.

\vskip6mm

We gratefully acknowledge J. Soto
for helpful discussions.
This work has been performed under Grants No. BFM2002-01868 from
DGESIC (Spain),
and No. 2001GR-0064
from Generalitat de Catalunya.

\eject

\end{document}